# Low temperature specific heat in BaFe$_{1.9}$Ni$_{0.1}$As$_2$ single crystals

ZENG Bin, MU Gang, LUO Huiqian & WEN Hai-Hu[†]

National Laboratory for Superconductivity, Institute of Physics, Chinese Academy of Sciences, Beijing 100190, China; Beijing National Laboratory for Condensed Matter Physics, Institute of Physics, Chinese Academy of Sciences, Beijing 100190, China

Low-temperature specific heat was measured on the BaFe$_{1.9}$Ni$_{0.1}$As$_2$ single crystals with critical transition temperature $T_c$ = 20.1 K. A clear specific heat jump with the value $\Delta C/T|_{Tc} \approx$ 23 mJ/mol K$^2$ has been observed. In addition, a roughly linear magnetic field dependence of the electronic specific heat coefficient $\Delta\gamma$(H) is found in the zero-temperature limit, suggesting that at least one Fermi pocket, probably the hole derivative one, is fully gapped with a small anisotropy in the present sample. A slight curvature of the curve $\Delta\gamma$(H) may suggest a complex gap structure (anisotropic gap or nodes) at other Fermi surfaces.

Specific heat, BaFe$_{1.9}$Ni$_{0.1}$As$_2$, nodeless gap

## 1 Introduction

The discovery of high-temperature superconductivity in the iron pnictides has stimulated enormous interests in the field of condensed matter physics[1]. Studies from different experiments have been carried out on the so-called 122 system because of the availability of sizable single crystals[2-7]. One of the key issues to study the superconductivity mechanism is the symmetry of the superconducting gap in this new family of high-$T_c$ superconductors. Theoretically the so-called s$^\pm$ model has been proposed, where the sign of the order parameters is opposite between the electron and hole pockets [8]. Some experimental results suggest nodeless superconducting gaps for the 122 system on both the hole-doped and electron-doped sides[2,4,5,6,9]. Meanwhile, multiple gaps with different anisotropies were observed in these experiments, indicating the complexity of the gap structure. Moreover, impurity scattering effect is considered to be important in understanding the experimental results with different pairing symmetries.[10,11,12]

For the Ni-doped BaFe$_{2-x}$Ni$_x$As$_2$, isotropic superconducting gaps with similar size on hole and electron pockets were suggested in optimal-doped sample from the magnetic field dependent thermal conductivity measurement[6]. However, the penetration depth measurement suggests a three-dimensional nodal superconducting gap in this system[13]. As for the specific heat measurements, only the specific heat jump near $T_c$ was reported in this Ni-doped 122 system[14]. So it is worthwhile to carry out the specific heat measurements under different fields to investigate the gap structure of this system. In this paper, we present the temperature and magnetic field dependent specific heat (SH) measurement on the optimal-doped BaFe$_{1.9}$Ni$_{0.1}$As$_2$ single crystals. A linear magnetic field dependence of the electronic specific heat coefficient $\Delta\gamma$(H) is found in the zero-temperature limit, suggesting that, at least one of the Fermi pocket, probably the hole doped one is fully gapped.

[†]Corresponding author (email: hhwen@aphy.iphy.ac.cn)
Supported by the National Natural Science Foundation of China (Grant Nos. 10221002/A0402 and 10774170/A0402), the National Basic Research Program of China (Grant Nos. 2006CB601000, 2006CB921107, and 2006CB921802), and the Chinese Academy of Sciences (ITSNEM)
Contributed by ZENG Bin, MU Gang, LUO Huiqian & WEN Hai-Hu

Citation: Zeng B, Mu G, Luo H Q, et al. Low temperature specific heat in BaFe$_{1.9}$Ni$_{0.1}$As$_2$ single crystals. Sci China Ser G, 2010, XXX: XXX, doi: XXX

## 2 Experiment

The Ni-doped $BaFe_{1.9}Ni_{0.1}As_2$ single crystals were grown by the self-flux method[15]. The samples for the present measurements have typical dimensions of 3×3×0.4 mm$^3$. The dc magnetization measurements were done with a superconducting quantum interference device (Quantum Design, SQUID). The resistivity was measured with an Oxford cryogenic system (Maglab-Exa-12) with the magnetic field up to 12 T. The specific heat were measured with a Quantum Design instrument physical property measurement system (PPMS) with the temperature down to 1.8 K and field up to 9 T. We employed the thermal relaxation technique to perform the specific heat measurements. To improve the resolution, we used a latest developed SH measuring frame from Quantum Design, which has negligible field dependence of the sensor of the thermometer on the chip as well as the thermal conductance of the thermal linking wires.

## 3 Results and discussion

In Figure 1, we show the temperature dependence of resistivity under various magnetic fields of $H$ = 0, 1, 3, 5, 7, 9, and 12 T, respectively. The middle temperature of the superconducting transition under zero field is found to be about 20.1 K. We also measured the dc magnetization of the sample with the zero-field cooling and field cooling processes with $H$ = 20 Oe, as shown in the inset (a) of Figure 1. The rather sharp transition of resistivity and dc magnetization suggests the high quality of our sample. In the inset (b) of Figure 1, we present the $H_{c2}$ vs. $T$ curve for this sample. The data are extracted from the resistivity data at the middle point of superconducting transition. Based on the single band model and assuming that the upper critical field is limited by the orbital pair breaking effect, we estimate the upper critical field at 0 K according to the Werthamer-Helfand-Hohenberg (WHH) formula[16] $H_{c2}$ = $-0.69(dH_{c2}/dT)|_{T_c} \times T_c$. Taking $T_c$ = 20.1 K, the calculated value of upper critical field for H//c is about 37 T. We should mention that the upper critical field in the iron pnictide superconductors in the low temperature limit may be governed by the spin Pauli limit effect, therefore the real value of the upper critical field at T = 0 K may change from this estimated one.

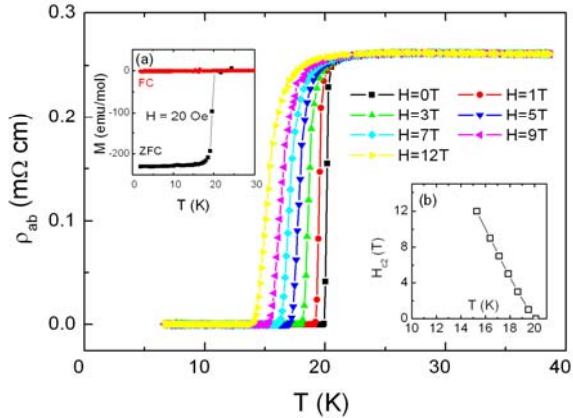

**Figure 1** Temperature dependence of resistivity for the sample $BaFe_{1.9}Ni_{0.1}As_2$ under different magnetic fields (H||c) is shown in the main frame. The inset (a) shows the temperature dependence of dc magnetization with the zero-field cooling and field cooling process with $H$ = 20 Oe. The inset (b) shows the phase diagram derived from the resistive transition curves defined by 50% $\rho_n$.

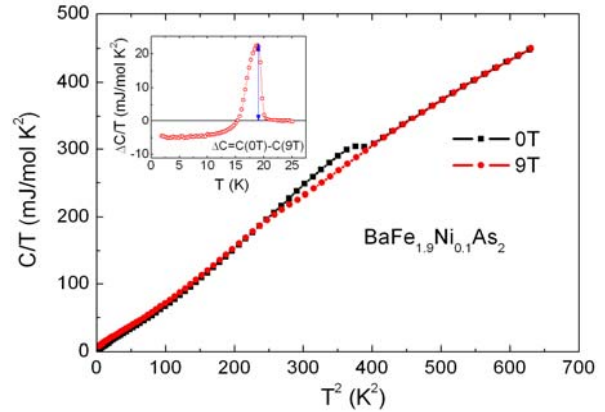

**Figure 2** Main frame: the raw data of specific heat for the sample $BaFe_{1.9}Ni_{0.1}As_2$ under 0 T and 9 T. Inset: the difference of specific heat between 0 T and 9 T. The arrowed blue line indicates the SH anomaly $\Delta C/T|_{Tc}$.

In order to have a comprehensive understanding, we measured the temperature and field dependent specific heat for this sample, as shown in Figure 2. Clear SH jump from superconducting transition can be seen in the raw data. We extracted the SH difference between 0 T and 9 T and showed the result in the inset of Figure 2. Although a magnetic field of 9 T does not suppress the superconductivity completely, it shifts the transition temperature to a distinguishable lower temperature. So the difference of C/T between 0 T and 9 T near $T_c$ can be taken as the SH anomaly $\Delta C/T|_{Tc}$. The value the SH anomaly evaluated from our data is about 23 mJ/mol K$^2$, which is similar to that reported by Bud'Ko S L et al[14], and a little smaller than that of the optimally doped Ba(Fe$_{1-x}$Co$_x$)$_2$As$_2$ sample.[11,14]

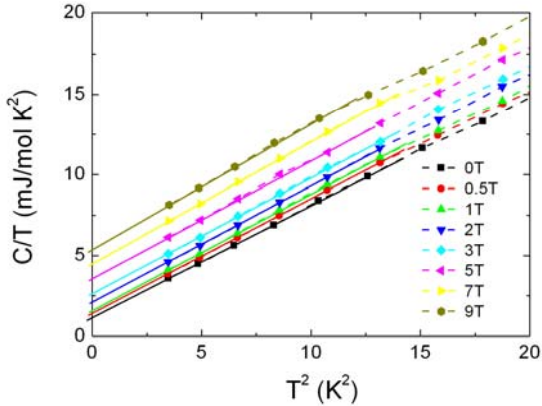

**Figure 3** Low-temperature specific heat plotted as $C/T$ vs. $T^2$ under different magnetic fields up to 9 T. No Schottky anomaly can be seen in the low temperature region. The solid lines present the linear extrapolations to zero temperature.

The raw data for the sample in the low-temperature region at different magnetic fields are plotted as C/T vs. $T^2$ in Figure 3. One can see the roughly linear behavior in the low-temperature region. No Schottky anomaly was detected in the sample; this may suggest that the Ni-doping induces no local paramagnetic centers, which would give a large contribution to SH as the Schottky anomaly in the low-temperature region. This is actually a very interesting point, since Ni doping, in many cases adds magnetic impurities into the system. In the iron pnictide system, the magnetic moment of these dopants, such as Ni and Co, may be well screened by the itinerant electrons. It is clear that the magnetic field enhances the low-temperature specific heat monotonicallyly, indicating the increase of quasiparticle DOS at Fermi level induced by magnetic field. In order to obtain the field induced term $\Delta\gamma(H) = [C(H)-C(0T)]/T$ at 0 K, we have extrapolated the SH data linearly to the zero-temperature limit, as shown by the solid lines in Figure 3.

The obtained field induced term $\Delta\gamma(H)$ at 0 K is shown in Figure 4. It can be seen clearly that $\Delta\gamma(H)$ increases almost linearly with the magnetic field. We all know that a single band superconductor with a nodal gap will exhibit a square-root behavior in the $\Delta\gamma(H)$ vs. $H$ plot because of the Doppler shift to the quasi-particle excitation spectrum induced by the supercurrents around the vortex cores[17]. So our data suggest that at least one FS pocket is fully gapped and the specific heat measures the contribution mostly from these bands. In the iron pnictide superconductors, the electrons in hole band is normally heavier than those in the electron band. Concerning that the specific heat is sensitive to the heavy electrons, we would suggest that the hole pocket is fully gapped. Actually, $\Delta\gamma$(H) vs. $H$ plot curve still has a weak non-linearity, which may suggest a complex gap structure at other Fermi surfaces. The present work actually reconciles the results from the thermal conductivity and penetration depth measurements[6,13].

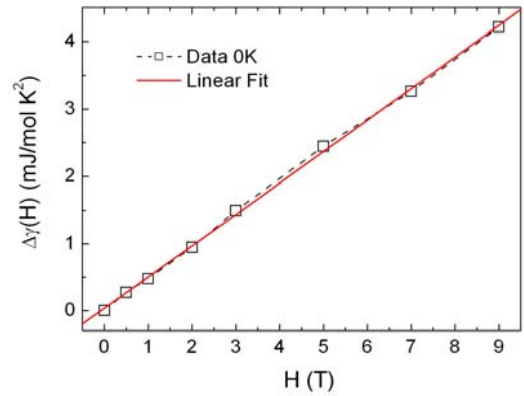

**Figure 4** The open squares represent the field dependence of the field-induced term $\Delta\gamma(H)$ at 0 K. The red line is a linear fit to the experimental data.

In a superconductor with nodeless gaps, it has been pointed out that $\Delta\gamma(H)$ is mainly contributed by the localized quasiparticle DOS within vortex cores[18]. As a result, one has $\Delta\gamma(H) = \gamma_n H/H_{c2}$, with $\gamma_n$ the normal state electron SH coefficient. From the linear fit in Figure 4, the slope $\Delta\gamma(H)/H$ is determined to be 0.47 mJ/(mol K$^2$T). Taking $H_{c2}$ = 37 T, $\gamma_n$ can be estimated to be about 17 mJ/mol K$^2$. This value is quite consistent with the specific heat anomaly $\Delta C/T|_{Tc}$= 23 mJ/mol K$^2$ with the scheme of weak coupling BCS: $\Delta C/T|_{Tc}$ = 1.43$\gamma_n$. From our data we get $\Delta C/T|_{Tc}/\gamma_n$=1.35.

## 4 Conclusions

In summary, we measured the resistivity and specific heat for the single crystal BaFe$_{1.9}$Ni$_{0.1}$As$_2$ under various magnetic fields. The superconducting transition tem-perature and upper critical field are determined to be 20.1 K and 37 T from the resistivity data, respectively. In the specific heat data, a clear SH anomaly with the value $\Delta C/T|_{Tc} \approx$ 23 mJ/mol K$^2$ is observed. The electronic specific heat coefficient $\Delta\gamma(H)$ in the zero-temperature limit shows a roughly linear dependence of the magnetic fields, suggesting that, at least one pocket, probably the hole pocket, is fully gapped. A slight curvature of the curve $\Delta\gamma(H)$ may suggest a complex gap structure at other Fermi surfaces. In addition, the normal state electron SH coefficient $\gamma_n$ is estimated to be about 17 mJ/mol K$^2$.

*The authors express their thanks to the people helping with this work, and acknowledge the valuable suggestions from the peer reviewers.*